\documentclass[9.5pt,conference]{IEEEtran}
\usepackage{amsmath}
\usepackage{amsthm}
\usepackage{cite}
\usepackage{amssymb}
\usepackage{graphicx}
\usepackage{algorithm}
\usepackage{algorithmic}
\usepackage{bm}

\ifCLASSINFOpdf
\else
\fi
\pdfoutput=1

\begin{document}

\title{Channel Estimation with Hierarchical Sparse Bayesian Learning for ODDM Systems}
\author{
	Jiasong~Han\textsuperscript{1},~Xuehan~Wang\textsuperscript{1},~Jingbo~Tan\textsuperscript{1},~Jintao~Wang\textsuperscript{1},~Yu~Zhang\textsuperscript{1},~Hai~Lin\textsuperscript{2},~Jinhong~Yuan\textsuperscript{3}\\
	\IEEEauthorblockA{
		\textsuperscript{1}Beijing National Research Center for Information Science and Technology (BNRist),\\
		Dept. of Electronic Engineering, Tsinghua University, Beijing, China\\
        \textsuperscript{2}Dept. of Electrical and Electronic Systems Engineering, Graduate School of Engineering,\\
        Osaka Metropolitan University, Sakai, Osaka 599-8531, Japan\\
        \textsuperscript{3}School of Electrical Engineering and Telecommunications, University of New South Wales, Sydney, NSW 2052, Australia\\
		\{hanjs22@mails., wang-xh21@mails., tanjingbo@, wangjintao@, zhang-yu@\}tsinghua.edu.cn, \\ hai.lin@ieee.org, j.yuan@unsw.edu.au}
\thanks{This work was supported in part by Tsinghua University-China Mobile Research Institute Joint Innovation Center.}	
}
\maketitle
\begin{abstract} 
Orthogonal delay-Doppler division multiplexing (ODDM) is a promising modulation technique for reliable communications in high-mobility scenarios. However, the existing channel estimation frameworks for ODDM systems cannot achieve both high accuracy and low complexity simultaneously, due to the inherent coupling of delay and Doppler parameters. To address this problem, a two-dimensional (2D) hierarchical sparse Bayesian learning (HSBL) based channel estimation framework is proposed in this paper. Specifically, we address the inherent coupling between delay and Doppler dimensions in ODDM by developing a partially-decoupled 2D sparse signal recovery (SSR) formulation on a virtual sampling grid defined in the delay-Doppler (DD) domain. With the help of the partially-decoupled formulation, the proposed 2D HSBL framework first performs low-complexity coarse on-grid 2D sparse Bayesian learning (SBL) estimation to identify potential channel paths. Then, high-resolution fine grids are constructed around these regions, where an off-grid 2D SBL estimation is applied to achieve accurate channel estimation. Simulation results demonstrate that the proposed framework achieves performance superior to conventional off-grid 2D SBL with significantly reduced computational complexity.
\end{abstract}


\IEEEpeerreviewmaketitle
\section{Introduction}
The next generation mobile communication systems require ultra-reliable communication in high-mobility scenarios such as high-speed railways (HSR), unmanned aerial vehicles (UAVs), and low earth orbit (LEO) satellites \cite{tan2}. Orthogonal frequency division multiplexing (OFDM) \cite{OFDM}, widely adopted in 4G and 5G networks, faces challenges in high-mobility scenarios due to Doppler-induced intercarrier interference (ICI), which leads to significant performance degradation \cite{OMP}.

To overcome this issue, orthogonal time frequency space (OTFS) modulation has been proposed in \cite{OTFS}. By modulating information in the delay-Doppler (DD) domain, OTFS exploits the quasi-time-invariant nature of doubly-selective channels, thereby outperforming OFDM under high mobility scenarios. However, OTFS suffers from high out-of-band emission (OOBE) caused by the discontinuity and rectangular pulse shaping \cite{OTFS_EP}, limiting its practical application. As a promising solution, orthogonal delay-Doppler division multiplexing (ODDM) modulation has been introduced in \cite{ODDM, ODDM_HL}. ODDM employs a pulse train designed to preserve orthogonality with respect to DD resolutions, demonstrating enhanced performance compared to OTFS.

The DD domain channel estimation plays an important role in ODDM and all the other DD domain modulation schemes, which has been investigated in many literatures \cite{pilot, OMP_W, SBL_conf, SBL_OTFS, tan1}. An impulse-based DD domain channel estimation method was proposed in \cite{pilot}. While, the simple thresholding operation in this method severely limits its performance. \cite{OMP_W} proposed an on-grid orthogonal matching pursuit (OMP) based channel estimation method to estimate the effective channel response. However, the on-grid delay and Doppler assumed in this method do not conform to the actual channel conditions, making it difficult to deal with the off-grid delay and Doppler in real channels. To this end, an one-dimensional (1D) off-grid sparse Bayesian learning (SBL) method was designed in \cite{SBL_conf} which can estimate the DD domain original channel response more accurately but suffers from high complexity. To address the complexity issue of the 1D SBL method, \cite{SBL_OTFS} transformed the channel estimation problem into a two-dimensional (2D) sparse signal recovery (SSR) formulation and then solved this problem by using a 2D SBL method, which significantly reduce the complexity. However, the 2D SSR formulation in \cite{SBL_OTFS} is established based on the bi-orthogonal assumption where the delay domain and Doppler domain are fully-decoupled and can be estimated separately. This ideal bi-orthogonal assumption does not exist in practical DD domain modulation such as ODDM \cite{ODDM_general}, where the coupling between the delay and Doppler domain cannot be separated, which makes the 2D SBL approach in \cite{SBL_OTFS} face performance degradation in practical systems.

To address the aforementioned issues on existing DD domain channel estimation methods and realize accurate channel estimation with low-complexity in ODDM systems, we propose a 2D hierarchical sparse Bayesian learning (HSBL) framework. A partially-decoupled 2D SSR formulation is first proposed to properly accommodate the coupled delay-Doppler structure in ODDM systems. Building upon this adapted formulation, we further develop a coarse-to-fine grid refinement framework that significantly reduces computational complexity. The method begins with a low-complexity coarse on-grid 2D SBL estimation to identify potential channel path locations, followed by targeted high-resolution fine-grid estimation only in these potential regions using off-grid 2D SBL. Simulation results demonstrate that the proposed approach not only effectively addresses the coupling challenge in ODDM channel estimation but also achieves superior performance with substantially reduced computational complexity compared to conventional off-grid 2D SBL method.
  

\section{System Model}
In this section, we review the basic concepts of ODDM and the corresponding input-output (IO) relation. We consider ODDM systems over general physical channels operating with symbol interval $T_s$ and subcarrier spacing $\frac{1}{NT_s}$, where each ODDM frame consists of $M$ multi-carrier symbols and each symbol has $N$ subcarriers. An ODDM frame carries $MN$ digital symbols $\{{X[m,n]}|m=0,1, \cdots, M-1, n=0,1, \cdots, N-1\}$ in the DD domain, where $X[m,n]$ indicates the data component at the $n$-th ODDM subcarrier of the $m$-th multicarrier symbol. At the transmitter, the ODDM waveform without cyclic prefix (CP) can be represented as
\begin{align}
    \tilde{x}(t) = \sum_{m=0}^{M-1} \sum_{n=-N/2}^{N/2-1} X[m,n] u\left(t - mT_s\right) e^{j 2\pi \frac{n}{MN T_s} \left(t - mT_s\right)},
\end{align}
\noindent where \(u(t)\) is a pulse-train defined as \( u(t) = \sum\limits_{\dot{n}=0}^{N-1} a(t - \dot{n} MT_s)\), and the subpulse $a(t)$ is a square-root Nyquist pulse parameterized by its Nyquist interval $T_s$ and its duration \( T_a = 2QT_s\), where $Q$ is a positive integer and \( 2Q \ll M\).

To mitigate inter-frame interference (IFI) between adjacent ODDM frames, a CP is appended, whose length is \( D_{\text{CP}} < M \) and \(D_{\text{CP}} T_s\) is larger than the maximum equivalent delay spread. As indicated in \cite{ODDM,ODDM_HL}, the continuous-time signal can be \textit{approximately} obtained by applying the sample-wise pulse shaping as
\begin{align}
    x(t) \approx \sum_{k=-D_{\text{CP}}}^{MN-1} x\left[ [k]_{MN} \right] \, a\left(t - k \frac{T}{M}\right),
\end{align}
\noindent where \( x[k] = \frac{1}{\sqrt{N}} \sum\limits_{n=0}^{N-1} X\left[[k]_M,n\right] e^{j \frac{2\pi}{N} n \left\lfloor \frac{k}{M} \right\rfloor} \) is the form of \(X[m,n]\) after inverse discrete Fourier transform (IDFT) and vectorization, and $[\cdot]_M$ is the modulo operation.

When the transmission signal \( x(t) \) passes the time-varying multi-path channel, according to the sparsity of the DD domain channel \cite{OTFS}, the baseband received signal can be written as
\begin{align}
    y(t) = \sum\limits_{p=1}^P h_p x(t-\tau_p) e^{j2\pi \nu_p (t-\tau_p)} + z(t) ,
\end{align}
\noindent where \( P \) is the number of propagation paths, \( h_p \), \( \tau_p \), and \( \nu_p \) represent the complex gain, delay, and Doppler shift associated with the \( p \)-th path, and $z(t)$ denotes the additive noise. 

For ease of illustration, let $l_p \triangleq \frac{\tau_p}{T_s}$ and $k_p \triangleq \nu_p MNT_s$ denote the normalized delay and Doppler shifts, which are not necessarily integer. The receiver employs correlators based on $MN$ time-frequency shifted $u(t)$, which can be approximated by a combination of an $a(t)$-based matched filtering (MF), $MN$ times sampling, and an $N$-point discrete Fourier transform (DFT). Then, the digital samples \(y[k]\) can be written as
\begin{align}
    & \nonumber y[k] = \left[y(t) * a^*(-t)\right]\Big|_{t=kT_s}   \\
    =& \sum\limits_{d=0}^D \sum\limits_{p=1}^P h_p g\left[ (d-l_p)T_s \right] e^{j2 \frac{(k-l_p)k_p}{MN}} x[[k-d]_{MN}] + z[k],
\end{align}
\noindent where $D$ is the total number of delay taps of the equivalent sampled channel and $g(\tau) \triangleq a(\tau) * a^*(-\tau)$ \cite{ODDM_general}.
\begin{figure*}[t]
	\begin{equation}
		\begin{aligned}
            Y[m,n] = \sum\limits_{p=1}^P \sum\limits_{d=0}^{D-1} \sum\limits_{\tilde{n}=0}^{N-1} \tilde{h}_p g \left[(d-l_p)T_s \right]e^{j2\pi \frac{mk_p}{MN}} \frac{1}{N} \frac{1-e^{-j2\pi (n-\tilde{n}-k_p)}}{1-e^{-j\frac{2\pi}{N} (n-\tilde{n}-k_p)}} \Psi[m, d, \tilde{n}] X[[m-d]_M, \tilde{n}] + Z[m,n]
		\end{aligned}
		\label{ddio}
	\end{equation}
	\hrulefill
\end{figure*}
Define the DD domain output as \(Y[m,n] = \frac{1}{\sqrt{N}} \sum_{\dot{n}=0}^{N-1} y[\dot{n}M + m] e^{-j 2 \pi \frac{\dot{n} n}{N}}\). On this basis, the corresponding input-output relationship is obtained as (\ref{ddio}) at the top of next page, where \(Z[m,n]\) denotes the noise in the DD domain, \( \tilde{h}_p = h_p e^{-j2\pi \frac{l_p k_p}{MN}} \) and 
\begin{align}
    \Psi[m, d, \tilde{n}] = 
    \begin{cases} 
    1, & 0 \leq d \leq m \leq M-1, \\
    e^{-j 2 \pi \frac{\widetilde{n}}{N}}, & 0 \leq m < d \leq D-1.
    \end{cases}
\end{align}


\section{Partially-Decoupled 2D SSR Formulation for ODDM Channel Estimation}
In this section, we provide the partially-decoupled 2D SSR formulation for ODDM systems. For notational simplicity, we consider a single pilot case. A pilot symbol is inserted at \( X[m_0,n_0]\) with a value of \( X_0 \). Similar to \cite{pilot}, the DD domain grids in the range $m_0 - D \leq m \leq m_0 + D$ and $0 \leq n \leq N-1$, with $m \neq m_0$ and $n \neq n_0$, remain null serving as guard space to mitigate interferences from data symbols to pilot symbols. The remained DD domain grids can be used for transmitting data symbols. \footnote{The extension to multiple pilots is essentially the same but with more complex expressions. Hence, we proceed without further elaboration.}

In particular, we only employ the received symbols which are affected by the pilot symbol. As such, the received signal in (\ref{ddio}) can be rewritten as
\begin{align}\label{element}
    &\nonumber Y[m,n] = X_0 \sum\limits_{p=1}^P \tilde{h}_p w_{\tau}(m, l_p, k_p) w_{\nu}(n, k_p) + Z[m,n], \\
    & m_0 \leq m \leq m_0 + D, \quad 0 \leq n \leq N-1,
\end{align}
\noindent where we may assume \( m_0 \leq M-D-1 \) without loss of generality, in which case \( \Psi[m, d, \tilde{n}] = 1\), and
\begin{align}
     w_{\tau}(m, l_p, k_p) &= g \left[(m-m_0-l_p)T_s \right]e^{j2\pi \frac{mk_p}{MN}}, \\
     w_{\nu}(n, k_p) &= \frac{1}{N} \frac{1-e^{-j2\pi (n-n_0-k_p)}}{1-e^{-j\frac{2\pi}{N} (n-n_0-k_p)}}.
\end{align}

Following the principle of compressed channel sensing \cite{SBL}, we formulate a partially decoupled 2D SSR problem based on a virtual sampling grid in the DD domain. A virtual sampling grid is defined over the ranges $(0, l_{\max})$ for the delay domain and $(-k_{\max}, k_{\max})$ for the Doppler domain, where \(l_{max}\) and \(k_{max}\) are the maximal normalized delay and Doppler shift of the channel, respectively. First, $N_0$ grid points are established in the Doppler domain, forming a vector $\mathbf{k} = [k_0, \ldots, k_{N_0-1}]^T \in \mathbb{C}^{N_{0} \times 1}$. For each Doppler grid point $k_i$ where $i \in \{0, \ldots, N_0-1\}$, $M_0$ grid points are defined in the delay domain, collectively forming a matrix $\mathbf{L} \in \mathbb{C}^{M_0 \times N_0}$ where each column corresponds to the delay grid points for a specific Doppler frequency. 

Considering an equally spaced sampling grid with virtual Doppler resolution $r_{\nu} = \frac{2k_{\max}}{N_0}$ and virtual delay resolution $r_{\tau} = \frac{l_{\max}}{M_0}$, we have \( k_i = i \cdot r_{\nu} - k_{\max} \) and \( l_{j,i} = j \cdot r_{\tau} \), \( \forall i \in \{0, \ldots, N_0-1\} \) and \( \forall j \in \{0, \ldots, M_0-1\} \), where $l_{j,i}$ represents the element at row $j$ and column $i$ of matrix $\mathbf{L}$. \( \Delta \mathbf{k} \in \mathbb{C}^{N_{0} \times 1}\) and \( \Delta \mathbf{L} \in \mathbb{C}^{M_0 \times N_0}\) are defined to represent the corresponding off-grid components of the Doppler and delay grid, respectively. Let \(\Delta k_i\) and \(\Delta l_{j,i}\) denote the $i$-th element of vector \(\Delta \mathbf{k}\) and the \((j,i)\)-th element of matrix \(\Delta \mathbf{L}\), respectively. Their values are confined to the intervals \(\Delta k_i \in \left[ -\frac{1}{2} r_{\nu}, \frac{1}{2} r_{\nu} \right] \) and \(\Delta l_{j,i} \in \left[ -\frac{1}{2} r_{\tau}, \frac{1}{2} r_{\tau} \right] \).

Based on (\ref{element}), the first-order approximation \cite{SBL} of the 2D SSR formulation could be written as
\begin{align}\label{ssr_element}
    \nonumber Y[m,n] \approx &\sum\limits_{i=0}^{N_0-1} \left[ \sum\limits_{j=0}^{M_0-1} \tilde{h}_{i,j} \overline{w_{\tau}}(m, l_{j,i}, k_i) \right] \overline{w_{\nu}}(n, k_i) \\
    & + Z[m,n],
\end{align}
\noindent where we assume \( X_0 = 1 \) and
\begin{align}
    \overline{w_{\nu}}(n, k_i) &= w_{\nu}(n, k_i) + \frac{\partial w_{\nu}}{\partial k_i}(n, k_i) \Delta k_i, \\
    \overline{w_{\tau}}(m, l_{j,i}, k_i) &= w_{\tau}(m, l_{j,i}, k_i) + \frac{\partial w_{\tau}}{\partial l_{j,i}}(m, l_{j,i}, k_i) \Delta l_{j,i},
\end{align}
\noindent and \( \tilde{h}_{i,j} \) represents the complex channel gain coefficient at the $(i,j)$-th sampling grid point, where $i$ and $j$ denote the indices in the Doppler and delay dimensions, respectively. If the grid point \( (k_{i}, l_{j,i}) \) is the closest to the delay and Doppler shift \( (k_p, l_p) \) of the $p$-th subpath, then \( \tilde{h}_{i,j} = \tilde{h}_p \), \(k_p = k_i + \Delta k_i\) and \(l_p = l_{j,i} + \Delta l_{j,i}\). Otherwise, \(\tilde{h}_{i,j}\) is zero.

We define a specialized tensor-matrix operation denoted by $\circledast$ that transforms a 3D tensor $\mathcal{X} \in \mathbb{C}^{A \times B \times C}$ and a matrix $\mathbf{Y} \in \mathbb{C}^{B \times C}$ into a matrix $\mathbf{Z} \in \mathbb{C}^{A \times C}$. This operation performs $C$ independent matrix-vector multiplications, where for each $k \in \{1, \ldots, C\}$, the $k$-th frontal slice of $\mathcal{X}$ ($\mathcal{X}_{:,:,k}$) is multiplied by the $k$-th column vector of $\mathbf{Y}$ ($\mathbf{Y}_{:,k}$). Formally, the operation is defined element-wise as \( \mathbf{Z}_{i,k} = \sum_{j=1}^{B} \mathcal{X}_{i,j,k} \cdot \mathbf{Y}_{j,k} \), \( \forall i \in \{ 1, \cdots A \} \) and \( \forall k \in \{ 1, \cdots C \} \).


Then the matrix form of (\ref{ssr_element}) could be written as
\begin{align}\label{model}
    \mathbf{Y} = \mathbf{K}(\Delta \mathbf{k}) \times \left[\mathcal{L}(\Delta \mathbf{L}) \circledast \mathbf{H}\right]^T + \mathbf{Z},
\end{align}
\noindent where \( \mathbf{Y} \in \mathbb{C}^{N \times D}\), \( \mathbf{Z} \in \mathbb{C}^{N \times D}\), \( {\mathbf{Y}}_{n,m} = Y[m+m_0,n]\), and \( \mathbf{Z}_{n,m} = Z[m+m_0,n]\). The Doppler matrix \( \mathbf{K}(\Delta \mathbf{k}) \in \mathbb{C}^{N \times N_0}\), the delay sensor \( \mathcal{L}(\Delta \mathbf{L}) \in \mathbb{C}^{D \times M_0 \times N_0}\) and the sparse coefficient matrix \( \mathbf{H} \in \mathbb{C}^{M_0 \times N_0}\) are give by \( \mathbf{K}(\Delta \mathbf{k})[n,i] = \overline{w_{\nu}}(n, k_i) \), \( \mathcal{L}(\Delta \mathbf{L})[m, j, i] = \overline{w_{\tau}}(m, l_{j,i}, k_i) \) and \( \mathbf{H}[i,j] = \tilde{h}_{i,j} \), respectively.


Due to the channel sparsity, $\mathbf{H}$ is $P$-sparse, making (\ref{model}) a 2D SSR problem. It is worth emphasizing that, different from the common fully-decoupled 2D SSR formulation in \cite{SBL_OTFS} based on the idea bi-orthogonal assumption, the derived 2D SSR formulation in (\ref{model}) enjoys a partially decoupled structure which describes the practical relationship between the delay and Doppler components in ODDM. This enables a more precise characterization of the channel estimation formulation. By utilizing the partially-decoupled structure, new channel estimation framework can be designed and performance improvements to the original 2D SBL algorithms \cite{SBL_OTFS} can be expected. To this end, we propose a 2D HSBL based channel estimation framework in the following section.

\section{Proposed Channel Estimation Algorithm}
In this section, we first derive both on-grid and off-grid 2D SBL estimation algorithms based on the established partially-decoupled 2D SSR formulation, addressing the limitation of existing methods which rely on a decoupling assumption and are confined to bi-orthogonal OTFS systems \cite{SBL_OTFS}. Building upon these algorithms, we subsequently propose the 2D HSBL based channel estimation framework. Finally, we analyze the complexity of these algorithms.

\subsection{On-Grid \& Off-Grid 2D SBL Algorithm}
We first examine the off-grid case and subsequently demonstrate that the on-grid formulation constitutes a special case of the off-grid framework. 

The off-grid 2D SBL algorithm based on the derived partially-decoupled 2D SSR formulation consists of two steps, as illustrated in \textbf{Algorithm \ref{alg:sbl}}. The first step estimates the composite matrix \( \mathbf{V} =  \mathcal{L}(\Delta \mathbf{L}) \circledast \mathbf{H}\) and the second step aims to recover \( \mathbf{H} \) based on the estimated \( \mathbf{V} \).

\subsubsection{First Step}
The partially-decoupled formulation (\ref{model}) can be rewritten as
\begin{align}
    \mathbf{Y} = \mathbf{K}(\Delta \mathbf{k}) \mathbf{U} + \mathbf{Z}, \label{SBL_IO}
\end{align}
\noindent where \( \mathbf{U} = \mathbf{V}^T \). \( \mathbf{Z} \) is a complex Gaussian noise with \( \beta_0 = \frac{1}{\sigma^2} \) and \( \sigma^2 \) being the noise variance. The distribution of \( \mathbf{Y} \) is therefore described by the probability distribution function of \( p(\mathbf{Y}|\mathbf{U};\Delta \mathbf{k}, \beta_0) = \prod\limits_{m=0}^{D-1} \mathcal{CN}(\mathbf{Y}[:,m]|\mathbf{K}(\Delta \mathbf{k}) \mathbf{U}[:,m], \beta_0^{-1}\mathbf{I}) \), where \( \beta_0 \) adopts a Gamma hyper-prior as \( p(\beta_0|a,b) = \Gamma(\beta_0|a,b) \) \cite{SBL_OTFS}. Let \( \mathbf{U} \in \mathbb{C}^{N_0 \times D} \) follows a two-stage hierarchical prior as \( p(\mathbf{U}|\bm{\alpha}) = \prod\limits_{m=0}^{D-1} \mathcal{CN}(\mathbf{U}[:,m]|\mathbf{0}, \bm{\Lambda}) \), where \( \bm{\Lambda} = \text{diag}(\bm{\alpha}) \) denotes the covariance matrix, \( \bm{\alpha} = [\alpha_1, \cdots, \alpha_{N_0}]^T \), \( p(\bm{\alpha}|\rho) = \prod\limits_{n_0=0}^{N_0-1} \Gamma(\bm{\alpha}[n_0]|1, \frac{\rho}{2}) \). The unknown off-grid Doppler variables follow a uniform distribution within their bounds as \( \Delta k_i \sim \mathcal{U}\left[ -\frac{1}{2} r_{\nu}, \frac{1}{2} r_{\nu} \right]\).

As such, the Bayesian framework can be used to obtain the conditional posterior distribution \( p(\mathbf{U}|\mathbf{Y}; \bm{\alpha}, \Delta \mathbf{k}, \beta_0) = \prod\limits_{m=0}^{D-1} \mathcal{CN}(\mathbf{U}|\bm{\mu}[:,m], \bm{\Sigma}) \). Note that 2D SBL is an iterative algorithm, the expressions of the expectation \( \bm{\mu} \) and variance \( \bm{\Sigma} \) of \(\mathbf{U}\) can be obtained as
\begin{align}
    & \bm{\mu}^{(t+1)} = \beta_0^{(t)} \bm{\Sigma}^{(t+1)} \mathbf{K}^H \mathbf{Y}, \label{mean} \\
    \nonumber & \bm{\Sigma}^{(t+1)} = \left[ \beta_0^{(t)} \mathbf{K}^H \mathbf{K} + \left(\bm{\Lambda}^{(t)}\right)^{-1} \right]^{-1} \\
    &= \bm{\Lambda}^{(t)} - \bm{\Lambda}^{(t)} \mathbf{K}^H \left[ \left(\beta_0^{(t)}\right)^{-1} \mathbf{I} + \mathbf{K} \bm{\Lambda}^{(t)} \mathbf{K}^H \right]^{-1} \mathbf{K} \bm{\Lambda}^{(t)}, \label{var}
\end{align}
\noindent where $t$ is the number of iterations. Denoting \(\mathbf{K} = \mathbf{K}\left( \Delta \mathbf{k}^{(t)} \right)\) for simplicity. The hyper-parameters \( \bm{\alpha} \), \( \beta_0 \) and off-grid components in DD domain \( \Delta \mathbf{k} \) can be updated by
\begin{align}
    & \alpha_i^{(t+1)} = \frac{ \sqrt{D^2 + 4 \rho \left(\sum\limits_{m=0}^{D-1} |\bm{\mu}^{(t+1)}[i,m]|^2 + \bm{\Sigma}_{i,i}^{(t+1)} \right)} -D}{2 \rho}, \label{alpha} \\
    & \beta_0^{(t+1)} = \frac{2a-2+DN}{2b+||\mathbf{Y} - \mathbf{K}\bm{\mu}^{(t+1)}||^2 + M \text{tr}(\mathbf{K}\bm{\Sigma}^{(t+1)}\mathbf{K}^H)}, \label{beta} \\
    & \Delta \mathbf{k}^{(t+1)} = \arg\min_{\Delta \mathbf{k}}\left\{ \Delta \mathbf{k}^T \mathbf{A} \Delta \mathbf{k} - 2 \mathbf{b}^T \Delta \mathbf{k} \right\}, \label{k}
\end{align}
\noindent where the specific expression of \( \mathbf{A} \) and \( \mathbf{b} \) in (\ref{k}) are given in (\ref{A}) and (\ref{b}) at the top of next page, with \(\mathbf{K}_0[n,i] = w_{\nu}(n, k_i)\) and \( \mathbf{K}_1[n,i] = \frac{\partial w_{\nu}}{\partial k_i}(n, k_i) \Delta k_i^{(t)}\).
\begin{figure*}[t]
	\begin{equation}
		\begin{aligned}
            \mathbf{A} = \text{Re} \left\{ \mathbf{K}_1^H \mathbf{K}_1 \odot \left( \bm{\mu}^{(t+1)} \left(\bm{\mu}^{(t+1)}\right)^H + \bm{\Sigma}^{(t+1)} \right) \right\}, \label{A}
		\end{aligned}
	\end{equation}
    \hrulefill
    \begin{equation}
		\begin{aligned}
            \mathbf{b} = \frac{1}{D} \sum\limits_{m=0}^{D-1} \text{Re} \left\{ \text{diag} \left( \bm{\mu}^{(t+1)}[:,m] \right) \mathbf{K}_1^H \left( \mathbf{Y}[:,m] - \mathbf{K}_0 \bm{\mu}^{(t+1)}[:,m] \right) \right\}. \label{b}
		\end{aligned}
	\end{equation}
	\hrulefill
\end{figure*}
The stopping criteria is attained if \( \frac{\|\bm{\alpha}^{(t)} - \bm{\alpha}^{(t-1)}\|}{\|\bm{\alpha}^{(t-1)}\|} < \delta \) or the number of iterations reaches the maximum number \(T_{\max}\), where \(\delta\) is a settled tolerance.

\subsubsection{Second Step}
Based on the output of the first step $\mathbf{V}=(\bm{\mu}^{(t)})^T$, the channel estimation problem in the second step can be expressed as
\begin{equation}\label{SBL_IO2}
    \resizebox{0.87\linewidth}{!}{$
        \mathbf{V}[:,i] = \mathcal{L}[:,:,i](\Delta \mathbf{L}[:,i]) \mathbf{H}[:,i], \quad \forall i \in \{0, \cdots, N_0-1\},
    $}
\end{equation}
\noindent which can be solved by performing the procedure outlined in the first step for \( N_0 \) times, incorporating the corresponding \(k_i\) estimated in the first step each time. As the aforementioned details have been discussed, they will not be reiterated here.

If the off-grid variables \(\Delta\mathbf{k}\) and \(\Delta\mathbf{L}\) are fixed to zero and are not updated, we can obtain the on-grid 2D SBL algorithm, as illustrated in \textbf{Algorithm \ref{alg:sbl}}, which is a simplified case of the off-grid framework.

\begin{algorithm}
	\caption{On-Grid and Off-Grid 2D SBL Algorithm}
	\label{alg:sbl}
	\begin{algorithmic}[1]
		\REQUIRE Received signal $\mathbf{Y}$, maximum number of iterarions $T_{max}$, iteration counter $t$, root parameters $a$, $b$ and $\rho$.
		\STATE \textbf{First Step}
		\STATE Initialize $t=1$, and the hyperparameters $\beta^{(t)} = \frac{1}{\sigma^2}$, $\bm{\alpha}^{(t)} = \frac{1}{D} \sum\limits_{m=0}^{D-1} \left\| \mathbf{K}^H \mathbf{U}[:,m] \right\|$, $\Delta \mathbf{k}^{(t)} = \mathbf{0}$.
		\STATE \textbf{repeat} 
		\STATE \quad Update $\bm{\mu}^{(t)}$ and $\bm{\Sigma}^{(t)}$ via Eqs. \eqref{mean} and \eqref{var}.
		\STATE \quad Update $\bm{\alpha}^{(t)}$ and $\beta^{(t)}$ via Eqs. \eqref{alpha} and \eqref{beta}.
        \STATE \quad Update $\Delta \mathbf{k}^{(t)}$ via Eq \eqref{k} for off-grid 2D SBL.
        \STATE \quad $t = t+1$.
		\STATE \textbf{until} stopping criteria.
        \STATE Output $\mathbf{V}=\left( \bm{\mu}^{(t)} \right)^T$, $\hat{\mathbf{k}} = \mathbf{k}$ for on-grid 2D SBL or $\mathbf{V}=\left( \bm{\mu}^{(t)} \right)^T$, $\hat{\mathbf{k}} = \mathbf{k} + \Delta \mathbf{k}^{(t)}$  for off-grid 2D SBL.
		\STATE \textbf{Second Step}
        \FOR{$i=0:N_0-1$}
        \STATE Extract estimated Doppler shift $\hat{k}_i$ from $\hat{\mathbf{k}}$.
            \STATE Apply SBL procedure (similar to First Step) to solve $\mathbf{V}[:,i] = \mathcal{L}[:,:,i](\Delta \mathbf{L}[:,i]) \mathbf{H}[:,i]$ using $\hat{k}_i$.
        \ENDFOR
        \STATE \textbf{return} $\hat{\mathbf{H}}$, $\hat{\mathbf{k}}$ and $\hat{\mathbf{L}} = \mathbf{L}$ for on-grid 2D SBL or $\hat{\mathbf{H}}$, $\hat{\mathbf{k}}$ and $\hat{\mathbf{L}} = \mathbf{L} + \Delta \mathbf{L}^{(t)}$ for off-grid 2D SBL.
	\end{algorithmic}		
\end{algorithm}

\subsection{2D Hierarchical SBL Algorithm}
Although the aforementioned SBL algorithms can effectively solve problem (\ref{model}), uniform sampling of the DD domain results in prohibitively high complexity. To reduce the channel estimation complexity, we propose a 2D HSBL channel estimation framework, as summarized in \textbf{Algorithm \ref{alg:hsbl}}, building upon the above SBL estimation algorithms. The core idea of 2D HSBL is to replace the exhaustive high-resolution search over the entire DD domain with a coarse-to-fine two-stage strategy, thereby confining the computationally expensive off-grid operations to a small number of likely regions.

\subsubsection{Stage I}
A coarse virtual sampling grid is constructed with significantly fewer points \( N_c < N_0 \) and \( M_c < M_0 \). This defines a grid with resolutions \( r_{\nu}^{(c)} = \frac{2k_{\max}}{N_c} \) and \( r_{\tau}^{(c)} = \frac{l_{\max}}{M_c} \). An on-grid 2D SBL estimation is then performed on this coarse grid. After convergence, the estimates of the channel gains can be obtained as \( \hat{\mathbf{H}}_c \). 

Define a set \( \mathcal{I} \) including the grid points whose estimated gains exceed a threshold \( \eta \) as
\begin{align}
    \mathcal{I} = \left\{ (k_i,l_{j,i}) \left| \left| \hat{\mathbf{H}}_c[i, j] \right|^2 > \eta \cdot \max_{\forall i_0, j_0} \left| \hat{\mathbf{H}}_c[i_0, j_0] \right|^2 \right. \right\},
\end{align}
\noindent where \( \eta \in (0, 1) \) is a pre-defined threshold ratio. This set \( \mathcal{I} \) corresponds to the regions in the DD domain where true paths are likely to exist.

\subsubsection{Stage II}
Define a high-resolution search window centered at each significant coarse grid point \( \left(k_i, l_{j,i}\right) \). The Doppler and delay resolutions within this window are the desired fine resolutions \( r_{\nu}^{(f)} = \frac{2k_{max}}{N_f} \) and \( r_{\tau}^{(f)} = \frac{l_{max}}{M_f} \). The composite fine grid \( \mathcal{G}_{f} \) is formally defined as the union of all local fine grids constructed around each significant coarse grid point as
\begin{align}
\mathcal{G}_{f} = \bigcup_{(k_i,l_{j,i}) \in \mathcal{I}} \left\{ (k,l) \left| \begin{array}{l}
k = k_i + a \cdot r_\nu^{(f)}, \\
l = l_{j,i} + b \cdot r_\tau^{(f)}, \\
a,b \in \{-W, \cdots, W\}, \\
|k| \leq k_{\max}, \; 0 \leq l \leq l_{\max}
\end{array} \right. \right\},
\end{align}
\noindent where \( W \) defines the window size in grid points.

The union of all these local fine windows forms a new, highly reduced, virtual sampling grid. An off-grid 2D SBL algorithm is applied only on this reduced grid. The result of this second stage is the final estimate of the sparse channel matrix \( \hat{\mathbf{H}} \) and the corresponding off-grid parameters.
\begin{algorithm}[t]
	\caption{2D Hierarchical SBL (HSBL) Algorithm}
	\label{alg:hsbl}
	\begin{algorithmic}[1]
		\REQUIRE Received signal $\mathbf{Y}$, grid resolutions $r_{\nu}^{(c)}, r_{\tau}^{(c)}$ (coarse), $r_{\nu}^{(f)}, r_{\tau}^{(f)}$ (fine), window size $W$, threshold $\eta$.
		\STATE \textbf{Stage I: Coarse On-Grid Estimation}
		\STATE Construct coarse grid $\mathbf{k}_c$, $\mathbf{L}_c$ with resolutions $r_{\nu}^{(c)}$, $r_{\tau}^{(c)}$.
		\STATE Estimate $\mathbf{H}_c$ via on-grid 2D SBL according to Alg.  \ref{alg:sbl}.
		\STATE Identify significant point set $\mathcal{I}$ using threshold $\eta$.
		\STATE \textbf{Stage II: Fine Off-Grid Estimation}
		\STATE Construct composite fine grid $\mathbf{k}_f$, $\mathbf{L}_f$ by creating local windows of size $W$ around each point in $\mathcal{I}$ with resolutions $r_{\nu}^{(f)}$, $r_{\tau}^{(f)}$.
		\STATE Estimate $\mathbf{H}_f$, $\Delta \mathbf{k}_f$, $\Delta \mathbf{L}_f$ via off-grid 2D SBL according to Alg. \ref{alg:sbl}.
		\STATE \textbf{return} $\hat{\mathbf{H}} = \mathbf{H}_f$, $\hat{\mathbf{k}} = \mathbf{k}_f + \Delta \mathbf{k}_f$, $\hat{\mathbf{L}} = \mathbf{L}_f + \Delta \mathbf{L}_f$
	\end{algorithmic}		
\end{algorithm}

\subsection{Complexity Analysis}
Here, we analyze the comlexity of the proposed 2D HSBL framework. Firstly, the complexity of the 2D SBL algorithm in \textbf{Algorithm \ref{alg:sbl}} is provided. In each iteration, we first calculate Eqs. (\ref{mean}) and (\ref{var}) with the complexity order \( \mathcal{O}(NN_0^2)\). Updating the parameters \( \bm{\alpha} \) and \( \beta_0\) by Eqs. (\ref{alpha}) and (\ref{beta}) costs \( \mathcal{O}(NN_0^2) \). Calculating the off-grid parameters by Eq. (\ref{k}) costs \( \mathcal{O}(P^3) \) by discarding the negligible components as mentioned in \cite{SBL_OTFS}, where \(P\) is the number of paths and \(P \ll N_0\). As such, the complexity order of this step is \( \mathcal{O}(NN_0^2) \). Similarly, the complexity order of solving Eq. (\ref{SBL_IO2}) is \( \mathcal{O}(DM_0^2) \) for \( i = 1, \cdots, N_0 \), respectively. Thus, the complexity order of off-grid 2D SBL is \( \mathcal{O}(NN_0^2) + \mathcal{O}(DN_0M_0^2) = \mathcal{O}(DN_0M_0^2)\), since \(M_0 \approx N_0\). Despite the additional off-grid parameter update in off-grid 2D SBL, the overall complexity order remains dominated by the same higher-order term \( \mathcal{O}(DN_0M_0^2) \). Therefore, the complexity order of on-grid 2D SBL is approximately the same as that of off-grid 2D SBL.

The proposed 2D HSBL algorithm first performs a coarse-grid estimation using \(M_c\) delay grid points and \(N_c\) Doppler grid points. It then refines the estimation within localized regions, containing  \(M_f = (2W+1)\eta M_c\) delay points and \(N_f = (2W+1)\eta N_c\) Doppler points. Therefore, the complexity order of constructing fine grid is \( \mathcal{O}\left( W M_c N_c \right) \), leading to a total complexity order of \( \mathcal{O}\left( DN_c M_c^2\right) + \mathcal{O}\left( DN_f M_f^2\right) + \mathcal{O}\left( W M_c N_c \right) = \mathcal{O}\left( k DN_c M_c^2\right) \), where \(k = (2W + 1)^3 \eta^3 + 1\). The complexity order comparison between the proposed 2D HSBL algorithm and some benchmark algorithms are summarized in TABLE \ref{tab:1}. Under the parameter configuration employed in the next section's simulations (\( M_0 = M_f = 2.5M_c \), \( N_0 = N_f = N_c \), \( W=2 \), \( \eta=0.15 \)), it can be found that the 2D SBL algorithm in \cite{SBL_OTFS} imposes approximately an order of magnitude higher computational complexity than the proposed 2D HSBL.

\begin{table}[h!]
    \centering
    \caption{COMPLEXITY OF DIFFERENT ALGORITHMS}
    \renewcommand\arraystretch{1.5}
    \begin{tabular}{|p{15em}|p{10 em}|}
        \hline
        OMP \cite{OMP_W} & \( \mathcal{O}(PDNM_0N_0) \) \\
        \hline
         On-Grid \& Off-Grid 2D SBL \cite{SBL_OTFS} & \( \mathcal{O}(DN_0M_0^2) \) \\
        \hline
        2D HSBL & \( \mathcal{O}\left( k DN_c M_c^2 \right) \) \\
        \hline
    \end{tabular}
    \label{tab:1}
\end{table}

\section{Simulation Results}


In this section, we present the simulation results of the proposed channel estimation algorithm. The simulation parameters include a carrier frequency of $5$ GHz, $\frac{1}{NT_s} = 15 \text{kHz}$, ODDM frame size of $(256, 64)$, maximum speed of $500$ km/h, cyclic prefix length of $32$ and modulation of 4-QAM. The delay-power profile follows the Tapped Delay Line-C (TDL-C) model, as specified in \cite{3gpp_ts_25_221}. We define $E_b$ as the energy per bit and $N_0$ as the noise power per unit bandwidth. The ratio \( \frac{E_b}{N_0} \) is adopted as the signal-to-noise ratio (SNR) metric for the system. The power of the pilot symbol is \(30\) dB higher than that of the data symbols. In our proposed \textbf{Algorithm \ref{alg:hsbl}}, we set \( r_{\nu}^{(f)} = r_{\tau}^{(f)} = 0.2 \), \( r_{\nu}^{(c)} = r_{\tau}^{(c)} = 0.5 \), \(\eta = 0.15\), \(W=2\), \(\delta = 10^{-3}\), \( \rho = 10^{-2} \) and \(a=b=10^{-4}\). For all the other algorithms, we set \( r_{\nu} = r_{\tau} = 0.2 \) as comparison. The traditional threshold-based channel estimation and the OMP algorithm are described by \cite{pilot}, \cite{OMP_W}, respectively.

\begin{figure}
    \centering
    \includegraphics[width=0.75\linewidth]{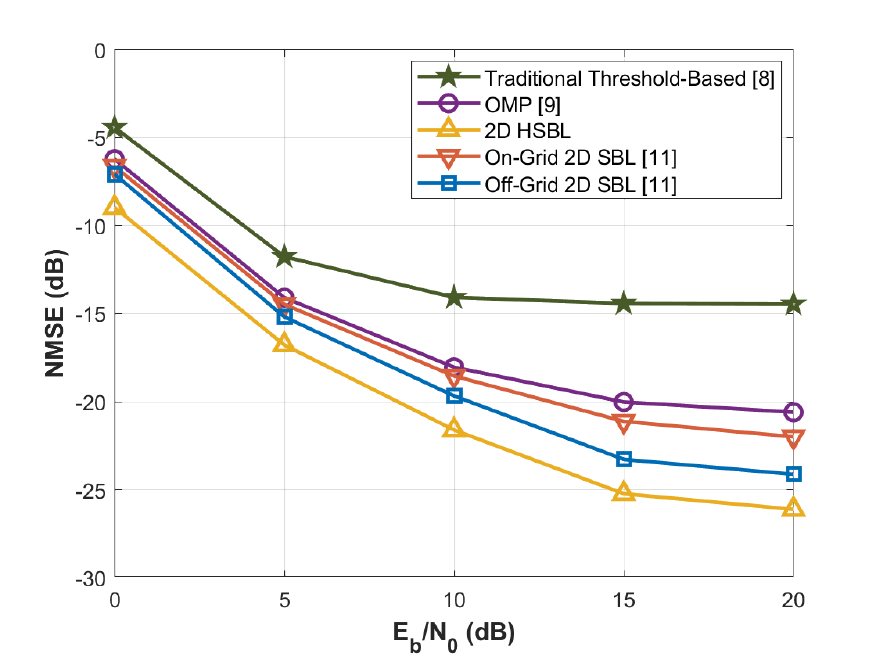}
    \caption{NMSE of channel estimation against SNR.}
    \label{nmse}
\end{figure}

\begin{figure}
    \centering
    \includegraphics[width=0.75\linewidth]{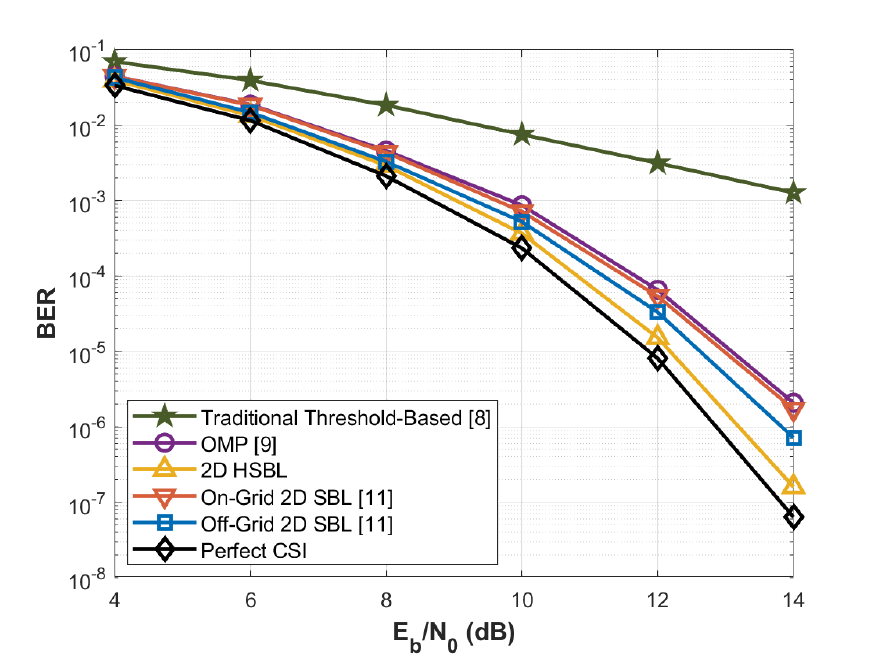}
    \caption{BER against SNR under 4-QAM alphabets and SIC-LMMSE detector.}
    \label{ber}
\end{figure}

Fig. \ref{nmse} illustrates the normalized mean squared error (NMSE) performance against SNR. The 2D HSBL algorithm achieves superior accuracy even compared to the off-grid 2D SBL, despite its significantly lower complexity. This enhancement stems from the hierarchical structure. The coarse estimation stage effectively eliminates irrelevant grid points, allowing the subsequent fine estimation to focus only on promising regions. In contrast, the off-grid 2D SBL estimates the entire grid, introducing collective noise from spurious non-zero entries, while 2D HSBL avoids for a cleaner reconstruction.

Fig. \ref{ber} illustrates the bit error rate (BER) performance under 4-QAM modulation versus SNR, where the SIC-LMMSE detection algorithm \cite{LMMSE} is utilized for signal detection. 2D HSBL achieves performance close to that of perfect CSI, reaching a BER of $10^{-6}$ at $32$ dB SNR, which outperforms the off-grid 2D SBL. Both OMP and on-grid 2D SBL incur an SNR loss of about 1 dB relative to 2D HSBL at a BER of $10^{-5}$. The conventional thresholds-based method exhibits a high error floor around $10^{-3}$.

\section{Conclusion}

To address the coupled delay-Doppler channel estimation challenge in ODDM systems, this paper developed a partially-decoupled 2D SSR formulation and proposed a 2D HSBL framework. The 2D HSBL employs a two-stage coarse-to-fine strategy: first, potential path locations are detected through a low-complexity coarse-grid estimation; subsequently, a super-resolution off-grid refinement is applied exclusively to these identified regions. Simulations demonstrate that 2D HSBL not only significantly reduces computational complexity but also achieves superior NMSE performance over conventional off-grid 2D SBL, by avoiding spurious estimates on irrelevant grid points. The proposed 2D HSBL offers an efficient and high-accuracy solution for ODDM channel estimation.

\appendices

\section*{Acknowledgment}
This work was supported in part by the Beijing Natural Science Foundation under Grants QY25034 and L242083, in part by the National Natural Science Foundation of China under Grant 62401315, and in part by the Japan Society for the Promotion of Science (JSPS) Grants-in-Aid for Scientific Research (KAKENHI) under Grant 22H01491.

\bibliographystyle{IEEEtran}
\bibliography{ref-sum}

\end{document}